**Effects of different feeding frequencies on growth, feed utilisation, digestive enzyme activities and plasma biochemistry of gilthead sea bream (*Sparus aurata*) fed with different fishmeal and fish oil dietary levels**

Serena Busti[a], Alessio Bonaldo[a], Francesco Dondi[a], Damiano Cavallini[a], Manuel Yúfera[b], Neda Gilannejad[b], Francisco Javier Moyano[c], Pier Paolo Gatta[a], Luca Parma[a]*

[a]Department of Veterinary Medical Sciences, University of Bologna, Via Tolara di Sopra 50, 40064 Ozzano Emilia, Bologna, Italy

[b]Instituto de Ciencias Marinas de Andalucía (ICMAN-CSIC), Apartado Oficial, 11519 Puerto Real, Cádiz, Spain

[c]Department of Biology and Geology, Universidad de Almería, La Cañada de San Urbano, 04120 Almería, Spain

*Corresponding author*: Luca Parma, Department of Veterinary Medical Sciences, University of Bologna, Viale Vespucci 2, 47042 Cesenatico, FC, Italy.

*Tel*.: +39 0547 338931; *Fax*: +39 0547 338941

E-mail address: luca.parma@unibo.it (L. Parma)

**Abstract**

In the context of Mediterranean aquaculture many efforts have been made in terms of reducing marine-derived ingredients in aquafeed formulation. On the other hand, little attention has been paid to the manipulation of feeding frequency at the on-growing phase, where the high costs related to feeding procedures and the optimisation of feed efficiency and fish health are key aspects for the economic and environmental sustainability of the production cycle. The effects of different feeding frequencies (F) (1F) one meal day$^{-1}$, (2F) two meals day$^{-1}$, (3F) three meals day$^{-1}$ on growth, digestive enzyme activity, feed digestibility and plasma biochemistry were studied in gilthead sea bream (*Sparus aurata*, L. 1758) fed with high (FM30/FO15, 30% fishmeal FM, 15% fish oil, FO) and low (FM10/FO3; 10% FM and 3% FO) FM and FO levels. Isonitrogenous and isolipidic extruded diets were fed to triplicate fish groups (initial weight: 88.3 ± 2.4 g) by a fixed ration over 109 days. No significant effects of feeding frequency on overall performance, feed efficiency and feed digestibility during the on-growing of gilthead sea bream fed high or low fishmeal and fish oil dietary level were observed. Pepsin activity showed an apparent decrease in fish receiving more than one meal a day which was not compensated by an increased production of alkaline proteases (either trypsin or chymotrypsin), particularly in fish fed on low FM. Although there were no effects on growth and feed utilisation at increasing feeding frequency, trypsin decreased significantly with an increasing number of meals only under low FMFO diet. Thus, it seemed that consecutive meals could have amplified the potential trypsin inhibitor effect of the vegetable meal-based diet adopted. Most of the plasma parameters related to nutritional and physiological conditions were not affected by feeding frequency, however an effect on electrolytes ($Na^+$, $Cl^-$), cortisol and creatinine was observed. The higher level of plasma creatinine detected in fish fed a single daily meal with high FMFO level seems to be within physiological values in relation to the higher protein efficiency observed with this diet. However, it will require further attention to exclude a possible overload of kidney functionality. According to the results, gilthead sea bream seems able to maximise feed utilisation regardless of the number of meals, and this could be a useful indicator for planning feeding activity at farm level to optimise growth of fish and costs of feeding procedures.

**Introduction**

The key performance indicators of Mediterranean aquaculture such as growth, feed utilisation and survival have not improved during the last decade despite the combined efforts of the scientific community and the aquafeed industry, which have led to significant advances concerning the nutritional requirement of its main target species: sea bream (*Sparus aurata*) and sea bass (*Dicentrarchus labrax*) (Parma et al., 2020). The intensification of production systems and their possible effects on stress and welfare or the less explored interaction between nutrition and feeding management may have contributed to this stagnation (Parma et al., 2020). In this context the improvement in feeding management during the grow-out phase, represents one of the solutions for increasing the economic profitability and competitiveness of the production sectors (Dias et al., 2010). Feed management includes several variables such as food size, feeding rate, delivery techniques, spatial distribution, feeding time and frequency (Xie et al., 2011; Kousoulaki et al., 2015). Among these, feeding frequency is one of the key factors that may contribute to maximising feed utilisation reducing waste and production costs (Tian et al., 2015). Based on the species-specific feeding habits, anatomy and physiology of the digestive tract, feeding frequency and timing may modulate the gut feed transit with possible consequences on digestion efficiency and nutrient retention (NRC, 2011; Gilannejad et al., 2019). Several studies have been conducted to determine the optimum feeding frequency in gilthead sea bream and other fish species (Lee et al., 2000; Wang et al., 2009; Enes et al., 2015; Tian et al., 2015; Zhao et al., 2016; Sun et al., 2016; Oh et al., 2018; Guo et al., 2018; Okomoda et al., 2019; Imsland et al., 2019; Gilannejad et al., 2019). Most of these studies have been performed on juvenile fish, which is consistent with the high metabolic and feeding demand at this stage where incorrect feeding frequency can have critical consequences on growth and survival (Lee et al., 2016). At this stage, different feeding frequencies were associated with growth, feeding behaviour, feed utilisation, digestive physiology, gastric evacuation, blood biochemistry, antioxidant status, immune response, fish welfare and water quality (Booth et al., 2008; Xie et al., 2011; Yúfera et al., 2014; Guo et al., 2018; Imsland et al., 2019; Pedrosa et al., 2019). However, little attention has been paid to the manipulation of feeding frequency during the on-growing phase, where the high costs related to feeding procedures and the optimisation of feed efficiency and fish health are key aspects for the economic and environmental sustainability of the production cycle. In gilthead sea bream, several thorough studies have addressed the digestive physiology of larvae and juveniles in relation to feeding rate, feeding timing, feeding frequency and circadian rhythms (Yúfera et al., 2014, 1995; Mata-Sotres et al., 2015; Ortiz-Monís et al., 2018). Recently Gilannejad et al. (2019) in a short-term experiment with early juveniles demonstrated for the first time that changes in the daily feeding time and frequency did not modify the feed transit rate and residence time in gilthead sea bream juveniles. The authors also found that very frequent diurnal feeding led to higher apparent digestibility, thus having an effect on nutrient utilisation. However, few studies on this topic have been performed in this species during the on-growing stage, particularly to assess whether optimal feeding frequency is consistent when fish are fed on diets formulated with low levels of fishmeal (FM) and fish oil (FO). Hence, the aim of the present study was to explore the effect of different feeding frequencies on feed efficiency, digestive enzyme activity and plasma biochemistry during the on-growing of gilthead sea bream fed low and high FM and FO dietary levels.

**Materials and methods**

*2.1 Experimental diets*

The experimental diets were produced via extrusion technology (pellet size = 4.0 mm) by SPAROS Lda (Portugal). Ingredients and proximate composition are presented in Table 1. The diets were formulated with high and low FM and FO dietary levels (FM30/FO15 and FM10/FO3; 30% FM, 15% FO and 10% FM and 3% FO, respectively) to contain 46% protein and 17% lipid. Diets were formulated including FM and a mixture of vegetable ingredients currently used for sea bream in aquafeed (Parma et al., 2016). Yttrium oxide was included at 100 mg kg$^{-1}$ to determine feed digestibility.

*2.2 Feeding frequency and rearing conditions*

The experiment was carried out at the Laboratory of Aquaculture, Department of Veterinary Medical Sciences of the University of Bologna (Cesenatico, Italy). Gilthead sea bream were obtained from an Italian fish farm (Orbetello, GR) and adapted to the laboratory facilities for 7 days before the beginning of the trial. Afterwards, 60 fish tank$^{-1}$ (initial weight: 88.3 ± 2.4 g) were randomly distributed in 18 square tanks (800 L capacity) with a conical base. Each diet was randomly assigned to triplicate fish groups under different feeding frequencies (F) (1, 2, 3 meals day$^{-1}$): (1F) one meal per day$^{-1}$ at 8.30 am, (2F) two meals per day$^{-1}$ at 08.30 and 16.00 and (3F) three meals per day$^{-1}$ at 08.30, 12.30 and 16.00 respectively, over 109 days. Each fish tank received the same total fixed daily ration of feed but supplied in 1, 2, and 3 meals according to the treatments. The feed amount was determined based on the feed ingested by group 1, which was fed a fixed ration close to satiation (~ 95% of satiation level). The fixed ration level was determined every 10 days by oversupplying the feed to 1F group according to the apparent satiation procedures usually employed under feeding trials as described in Bonvini et al. (2018a). In short, fish were overfed by automatic feeder during 1 h meal. The uneaten feed was subsequently trapped by a feed collector at the water output of the tanks, dried overnight at 105°C and the weight (after moisture conversion) was deducted from feed delivered to determine feed ingested. The amount of each meal supplied to 2F and 3F groups was respectively 1/2 and 1/3 of the amount supplied to 1F group. Keeping the daily ration identical, different feeding frequencies were chosen as the only variable. Each meal lasted 1 h, and in case of uneaten pellets, these were collected, dried overnight at 105°C, and weighed for overall calculation.

Tanks were supplied with natural seawater and connected to a closed recirculation aquaculture system (RAS) with an overall water volume of 22 m$^{-3}$. The RAS consisted of a mechanical sand filter (PTK 1200, Astralpool, Barcelona, Spain), ultraviolet lights (PE 25 mJ/cm$^2$: 32m$^3$ h$^{-1}$, Blaufish, Barcelona, Spain) and a biological filter (PTK 1200, Astralpool, Barcelona, Spain). The water flow rate within each tank was set at 100% exchange every hour, while the overall water renewal in the RAS was 5% daily. Mean water temperature was maintained at 24 ± 1.0 °C throughout the experiment; photoperiod was held constant at 12 h day through artificial light. The oxygen level was kept constant at 8.0 ± 1.0 mg L$^{-1}$ through a liquid oxygen system connected to a software controller (B&G Sinergia snc, Chioggia, Italy). Ammonia (total ammonia nitrogen ≤ 0.1 mg L$^{-1}$) and nitrite (≤ 0.2 mg L$^{-1}$) were spectrophotometrically monitored once a day (Spectroquant Nova 60, Merck, Lab business, Darmstadt, Germany). At the same time salinity (30 g L$^{-1}$) was measured by a salt refractometer (106 ATC) and sodium bicarbonate was added on a daily basis to keep pH at 7.8–8.0.

*2.3 Sampling*

All the fish in each tank were individually weighed at the beginning and at the end of the trial. At sampling fish were anaesthetized (100 mg L$^{-1}$) or euthanised (300 mg L$^{-1}$) by tricaine methanesulfonate MS-222 (Sigma-Aldrich). Carcass proximate composition was determined at the beginning and at the end of the trial. In the former case, one pooled sample of 10 fish was sampled to determine initial proximate composition while in the latter case one pooled sample of 5 fish tank$^{-1}$ was collected to determine final proximate composition. Furthermore, wet weight of viscera, liver and perivisceral fat was individually recorded for 5 fish per tank to determine viscerosomatic index (VSI) hepatosomatic index (HSI) and mesenteric fat index (MFI).

At the end of the trial at 5 hours post meal (hpm), 3 fish per tank (n=9/treatment) were sampled and dissected to obtain their entire gastrointestinal tract. Samples were stored at −80 °C and then freeze-dried for digestive enzyme activity analysis. In order to understand if consecutive meals could have had an impact on the digestive process we performed the analyses of digestive enzyme activity after the last meal of each feeding frequency regime.

At the same time, blood from 5 fish per tank (n=15 fish per diet treatment) was collected from the caudal vein. Blood samples were collected in less than 5 min to avoid increase in cortisol levels originated by manipulation (Molinero et al., 1997). Samples were then centrifuged (3000 x g, 10 min, 4°C) and plasma aliquots were stored at −80 °C until analysis (Bonvini et al., 2018b).

*2.4. Digestibility experiment*

At the end of the growth trial, the remaining groups of fish were used to determine the apparent digestibility coefficient (ADC) of dry matter, protein and energy, by the indirect method with diets containing yttrium oxide. Fish were fed according to the different feeding frequencies. After that at 8 h post-prandial fish were euthanised by overdose of anaesthetic and faeces were collected after fish dissection and stripping distal intestine.

Immediately after the collection, faeces were pooled for each tank and kept at −20 °C until analysis for yttrium, dry matter, protein, and energy content. ADC was calculated as follows:

ADC = 100*(1− (dietary Y2O2 level/ faecalY2O2 level))*((faecal nutrient or energy level/dietary nutrient or energy level)).

All experimental procedures were evaluated and approved by the Ethical-Scientific Committee for Animal Experimentation of the University of Bologna, in accordance with European directive 2010/63/UE on the protection of animals used for scientific purposes.

*2.5 Calculations*

The calculations for the determination of different performance parameters were as follows: specific growth rate (SGR) (% day$^{-1}$) = 100 * (ln FBW- ln IBW) / days (where FBW and IBW represent the final and the initial body weights, respectively); feed intake (FI) (g kg ABW$^{-1}$ day$^{-1}$)=((1000 ∗ total ingestion)/(ABW))/days)) (where average body weight, ABW=(IBW+FBW)/2; feed conversion ratio (FCR) = feed intake / weight gain; protein efficiency rate (PER) = (FBW – IBW) / protein intake; nutrient gain = (final carcass dry matter DM, protein or lipid content, g - initial carcass DM, protein or lipid content, g) / ABW, kg / days;  nutrient retention = DM, protein or lipid gain / DM, protein or lipid intake (where DM, protein or lipid intake = nutrient intake, g / ABW, kg / days. Viscerosomatic index (VSI) (%) = 100 ∗ (viscera weight/body weight). Hepatosomatic index (HSI) (%) = 100 ∗ (liver weight/body weight). Mesenteric fat index (MFI) (%) = (mesenteric fat weight/body weight).

*2.6 Proximate composition analysis*

Proximate composition was determined for diets and whole body of sampled fish. Faeces were analysed for dry matter, protein and energy content. Moisture content was obtained by weight loss after drying samples in an oven at 105 °C until a constant weight was achieved. Crude protein was analysed as total nitrogen (N) by using the Kjeldahl method and multiplying N by 6.25. Total lipids were performed according to Bligh and Dyer's (1959) extraction method. Ash content was determined by incineration to a constant weight in a muffle oven at 450 °C. Gross energy was determined by a calorimetric bomb (Adiabatic Calorimetric Bomb Parr 1261; PARR Instrument, IL, U.S.A). Yttrium oxide levels in diets and faeces were determined using Inductively Coupled Plasma-Atomic Emission Spectrometry (Perkin Elmer, MA, USA) according to Bonaldo et al. (2011).

*2.7 Digestive enzyme activity analyses*

Stomach and proximal intestine, including the pyloric caeca, of each individual were separately homogenised in distilled water (1:3 w/v), and were centrifuged at 4 °C, 13,000 g, for 10 min. Supernatants were stored at –20 °C until being processed. Using the stomach homogenate, pepsin activity was measured according to the methodology described in Anson (1938). In brief, 10 µL of the enzyme extract was diluted in 1 mL of 0.1 M HCl-glycine buffer (pH 2.0) containing 0.5 % bovine haemoglobin. The mixture was incubated during 20 min at room temperature. The reaction was terminated by adding 0.5 mL of 20 % trichloroacetic acid (TCA) and was cooled down at 4 °C during 15 min to facilitate precipitation. After centrifuging at 13,000 g for 15 min at 4 ºC, 200 µL of the supernatant was used to measure the absorbance at 280 nm. One unit of enzyme activity was defined as 1 µg tyrosine released per minute using a specific absorptivity of 0.008 ug$^{-1}$ cm$^{-1}$ at 280 nm.

In the proximal intestine homogenate, trypsin and chymotrypsin activity were measured using Nα-Benzoyl-DL-arginine 4-nitroanilide hydrochloride (BAPNA) and N-Glutaryl-L-phenylalanine p-nitroanilide (GAPNA) as substrates, according to Erlanger et al. (1961) and (1966) respectively. For each one of these enzymes, substrate stock (0.5 mM of BAPNA or GAPNA in dimethyl sulfoxide) was brought to the working concentration by 1/10th dilutions using 50 mM Tris-HCl and 20 mM $CaCl_2$ buffer (pH 8.5). The change in the absorbance at 405 nm was measured over 10 min, for 10 to 15 µL of the enzyme extract and 200 µL of substrate per each microplate well. For these enzymes, one unit of activity was defined as 1 µmol p-nitroaniline released per minute using coefficients of molar extinction of 8270 M$^{-1}$ cm$^{-1}$ at 405 nm.

Amylase activity was measured following the 3,5-di-nitrosalicylic acid (DNSA) method (Bernfeld, 1955). In brief, 30 µL of enzyme extract and 300 µL of substrate (2% soluble starch in 100 mM phosphate and 20 mM $NaCl_2$ buffer (pH 7.5) were incubated at 37 °C for 30 min. The reaction was stopped by the addition of 150 µL DNSA and was heated in boiling water for 5 min. After cooling on ice, 1.5 mL of distilled water was added to the mixture and the absorbance was measured at 530 nm. One unit of amylase activity was defined as the amount of enzyme needed to catalyse the formation of 1 µg of maltose equivalent per minute.

Lipase activity was measured using 4-Nitrophenyl myristate as substrate, according to Albro et al. (1985). Briefly, 10 µL of enzyme extract was added to 50 µL Sodium taurocholate (0.4 mg mL$^{-1}$) and 130 μL of 100 mM Tris-HCL buffer (pH 8.0) per each microplate well. The change in the absorbance at 405 nm was measured over 10 min. One unit of amylase activity was defined as the amount of enzyme needed to catalyse the production of 1 μg of p-nitrophenol per minute.

All the activities were expressed in units per g of wet weight of fish, considering both the total amount of tissue used for enzyme determination and the live weight of each sampled fish.

As previously indicated, in order to reduce fish stress to a minimum, only one sampling time collected after the last daily meal was used for measurement of digestive enzyme production. Assuming that a similar amount of enzyme should be released after each meal, total daily production of enzyme in treatments receiving two and three meals per day was estimated by multiplying the measured value by 2 or 3, respectively.

*2.8 Metabolic parameters in plasma*

Glucose (GLU), urea, creatinine, uric acid, total bilirubin (Tot bil), cholesterol (CHOL), high density lipoprotein (HDL), triglycerides (TRIG), total protein (TP), albumin (ALB), aspartate aminotransferase (AST), alkaline phosphatase (ALP), creatine kinase (CK), lactate dehydrogenase (LDH), calcium (Ca$^{+2}$), phosphorus (P), potassium (K$^+$) sodium (Na$^+$), iron (Fe), chloride (Cl), magnesium (Mg), and cortisol were measured in the plasma using samples of 500 μL on an automated analyser (AU 400; Beckman Coulter) according to the manufacturer's instructions. The ALB/globulin (GLOB), Na/K ratio and Ca x P were calculated.

*2.9 Statistical analysis*

All data are presented as mean ± standard deviation (SD). A tank was used as the experimental unit for analysing growth performance and feed digestibility, a pool of five sampled fish were considered the experimental unit for analysing carcass composition. Individual fish were used for analysing morphological parameters, digestive enzyme activity and plasma biochemistry. Data were analysed by a two-way analysis of variance (ANOVA) with Tukey's post hoc test. The normality and homogeneity of variance assumptions were validated for all data preceding ANOVA using Levene's Test for homogeneity of variance and Shapiro-Wilks normality test. Statistical analyses were performed using GraphPad Prism 6.0 for Windows (Graph Pad Software, San Diego, CA, USA). The differences among treatments were considered significant at $p \leq .05$.

**3. Results**

*3.1 Growth*

Results on growth performance parameters are summarised in Table 2. No significant effects on growth (FBW, weight gain and SGR) were detected between feeding frequencies for both dietary treatments ($p > .05$); however, fish fed FM30/FO15 displayed higher FBW, weight gain and SGR when compared to the FM10/FO3 groups (diet effect $p < .05$). Specifically, SGR of fish fed FM30/FO15$_{F1}$ was higher than those fed FM10/FO3$_{F1}$ and FM10/FO3$_{F3}$. FI was lower in FM30/FO15$_{F2}$ compared to FM10/FO3$_{F2}$ (diet effect $p < .05$). No significant effect of feeding frequencies on FCR was observed ($p$

> .05), however the FM10/FO3 group showed higher FCR values (diet effect $p < .05$) than FM30/FO15. There was no significant difference in survival rates among treatments ($p > .05$). Data on body composition and nutritional indices are shown in Table 3. Whole body composition values were not significantly affected by different feeding frequencies ($p > .05$), while protein content was lower in fish fed the FM10/FO3 diet; specifically FM10/FO3$_{F1}$ group was lower compared to the FM30/FO15$_{F2}$ and FM30/FO15$_{F3}$ groups ($p < .05$). Ash and moisture levels were not significantly affected by diet nor frequencies ($p > .05$), while a significant interaction was found for ash. No significant effects of feeding frequency on PER, nutrient gain and retention were detected ($p > .05$); however, fish fed FM10/FO3 displayed lower PER, DM and protein gain, and lower DM, protein and lipid retention compared to FM30/FO15 (diet effect $p < .05$).

No significant effect of feeding frequency was detected for HSI, VSI and MFI under both dietary treatments; however, MFI was higher ($p < .05$) in FM30/FO15 compared to FM10/FO3 (diet effect $p < .05$).

*3.2 Digestive enzyme activity*

Activity of the different digestive enzymes expressed per meal and estimated per day are shown in Figure 1 A-E and Fig. 1 a-e, respectively, while the ratio trypsin/chymotrypsin is shown in Fig 1F. Pepsin activity was significantly affected (frequency effect $p < .05$) by feeding frequencies with decreasing values from F1 to F2 in FM30/FO15, afterwards pepsin activity displayed similar level in F3 (Fig. 1A). When comparing different diets, pepsin activity was higher (diet effect $p < .05$) in FM30/FO15$_{F1}$ than FM10/FO3$_{F1}$. When analysing the estimated total release of enzyme per day, no significant differences were observed between feeding fish once or twice but the production of enzymes was significantly higher ($p < .05$) in those fed three times for both dietary treatments (Fig 1 a). The estimated total daily pepsin activity was also generally higher in FM30/FO15 than FM10/FO3 (diet effect $p = 0.045$) (Fig. 1a).

Trypsin activity was maintained constant ($p > .05$) irrespective of the number of meals in fish fed on FM30/FO15, while in those receiving the FM10/FO3 diet it was reduced from F2 to F3 (Fig 1B). The profile obtained when calculating the total activity released per day evidenced a significant increase of trypsin production with feeding frequency in fish fed on FM30/FO15, while in FM10/FO3 the activity did not show a significant increase (Fig 1b). Daily trypsin activities representation showed lower values (diet effect $p < .05$) in FM10/FO3$_{F3}$ compared to FM30/FO15$_{F3}$ (Fig 1b).

Feeding frequencies and dietary treatments did not affect chymotrypsin activity (Fig 1C). Similarly to trypsin, the estimated daily chymotrypsin activities showed an increase at increasing feeding frequency; specifically in FM30/FO15 a significant increase was observed between F2 and F3, while in FM10/FO3 no increase was evident between the same frequencies. Amylase activity displayed under both diets an increasing pattern between F1 and F2 followed by a decreasing trend in F3 (frequency effect $p < .05$). When total activity per day was calculated, activities increased from F1 to F2 under FM30/FO15 then remained constant in F3. When comparing different diets, amylase activity was generally lower (diet effect $p < .05$) in FM10/FO3 than in FM30/FO15 (Fig 1D, d) while within the same feeding frequency differences were more evident in F3 under daily estimation (Fig. 1d).

Lipase activity was not affected by feeding frequency nor diet (Fig. E). When estimating the total production per day a significant frequency effect was observed in particular under FM30/FO15. Trypsin/chymotrypsin was not affected by feeding frequency but resulted generally lower (diet effect $p = 0.048$) in FM10/FO3 compared to FM30/FO15.

*3.3 Feed digestibility*

Data on apparent feed digestibility are reported in Table 4. No significant differences were observed in dry matter, protein, and energy digestibility according to the different feeding frequencies. DM and protein digestibility was higher (diet effect $p < .05$) in FM10/FO3 compared to FM30/FO15.

*3.4 Plasma biochemistry*

The results of plasma parameters are shown in Table 5. Creatinine was affected ($p < .05$) by feeding frequency with higher values in FM30/FO15$_{F1}$ compared to FM30/FO15$_{F2}$ and FM30/FO15$_{F3}$. A significant interaction between frequency and diet was also observed. Na$^+$ was affected by feeding frequency with lower values in FM30/FO15$_{F1}$ compared to FM30/FO15$_{F2}$ and lower values in FM10/FO3$_{F1}$ compared to FM10/FO3$_{F2}$ and FM10/FO3$_{F3}$. A similar pattern was observed for Cl which was lower in F1 treatments compared to F2 and F3 under FM30/FO15 while under FM10/FO3 it was lower in F1*vs* F3. Fe was significantly higher in FM10/FO3 compared to FM30/FO15 (diet effect $p < .05$). Cortisol was affected by feeding frequency with an increasing pattern under F2 treatments in both diets. In addition cortisol was marginally (diet effect $p = 0.052$) higher in FM10/FO3 than FM30/FO15. No significant differences related to feeding frequency and diet were observed for Glucose, Urea, Uric acid, Tot bil, CHOL,

HDL, TRIG, TP, ALB, AST, ALP, CK, LDH, $Ca^{+2}$, P, $K^+$, Mg, ALB/GLOB, CaxP and Na/K while a significant interaction of the two effects was observed in Na/K

**Discussion**

The assessment of optimal feeding frequency plays a key role on overall performance, physiological responses and digestive efficiency in fish species, and although several studies have already investigated those effects on gilthead sea bream, the possible interaction between feeding frequency and different FM and FO dietary level at the on-growing stage of this species has been less explored. Previous studies on sea bream juveniles (17.9-21.9g) fed on a fixed ration but distributed into different numbers of meals have found that frequent diurnal feeding led to highest nutrient utilisation in comparison to a single feeding (Gilannejad et al., 2019). A similar result was reported in early juveniles of Asian sea bass (0.2-1.2g), which achieved maximum growth, survival and better feed conversion when a given ration was delivered three times a day (Biswas et al., 2010). Other studies also performed in juveniles of carnivorous species but fed to satiation have shown that growth and feed utilization generally increase with feeding frequency up to a given limit (Booth et al., 2008; Enes et al., 2015). The effect of feeding frequency on growth performance changes with the size of the fish. Generally, fry and juveniles are fed smaller meals more frequently, while feeding once or twice a day may be sufficient for broodstock and larger fish (Biswas et al., 2010). There may be several reasons for the decreases in feed efficiency associated with higher feeding frequencies: if intervals between meals are too short, the increase in gut transit rates results in less effective digestion and also, the increased swimming activity observed when feeding throughout long periods may lead to higher energy expenditure (Johansen and Jobling, 1998; Liu and Liao 1999; Tian et al., 2015). In the present study, results of FCR, nutritional indices and feed apparent digestibility did not reveal any effect of feeding frequencies on feed utilisation of either high and low FM and FO diets, indicating that the entire process of digestion and nutrient uptake was efficient regardless of the number of meals. In addition, the absence of a significant interaction indicated that the combined effect of frequency and diet did not alter the parameters studied. In particular, feeding frequency did not amplify or reduce the differences observed between diets. In the present study the fixed ration approach was chosen for its similarity to practical field application, but this has probably limited the maximum feed intake and growth which would have been reached in F2 and F3 treatments if an overfeeding regime had been adopted. However, the growth rate achieved was still surprising, especially if we consider that the daily feed amount provided in one meal led to an overall growth rate which is in line with previously reported studies using similar feed formulation, but conducted with an apparent satiation approach using more than one daily meal (Parma el al., 2020).

In agreement with the present study, Costa-Bomfim et al. (2014) did not find any significant improvement on SGR and FCR related to the provision of more than one daily meal in cobia at the on-growing stage (108-303g). This has also been described in tiger puffer (179-328g) where increasing from 1 up to 3 meals did not improve weight gain and FCR. Costa-Bomfim et al. (2014) hypothesized that a possible explanation for the lack of significance in cobia growth at an increasing number of daily meals could have been related to the food passing too rapidly through the digestive tract so that there may not be enough time to digest and absorb the feed efficiently. It has actually reported that feeding too frequently could lead to poorer feed conversion ratio due to "gastrointestinal overload", where intake of the next meal occurs before the previous bolus has been subjected to adequate gastric attack (Booth et al., 2008) and the existing chyme may enter the anterior intestine only partially digested (Riche et al., 2004). It is therefore clear that in order to understand how consecutive meals could have had an impact on the efficiency of the digestive process, several factors must be considered: a) if variations in gut transit rates substantially modify the time available for hydrolysis by digestive enzymes; b) if total enzyme production associated with a meal is enough to maintain a suitable enzyme:substrate ratio for effective hydrolysis; and c) if variations in gut transit rates affect the possibility of reaching pH values suitable for optimum functionality of the enzymes in the different gut sections. This last factor can be additionally influenced by the composition of the diet, more precisely by differences in the buffering capacity of some ingredients. In the present study it is presumed that fish fed repeatedly at intervals of 3.5 and 7.5 hours (F3 and F2, respectively) would be not able to complete the gastric digestion and evacuation of the previous meal in agreement with Nikolopoulou et al. (2011) who pointed out that 10 h would be required to complete gastric evacuation after a single meal in seabream (150±30g). This negative effect could be partially counteracted by variations in total enzyme production. In the present study it was observed that the amount of pepsin produced after a meal was influenced by the number of meals, showing an apparent decrease in fish receiving more than one meal a day that was not compensated by an increased production of alkaline proteases (either trypsin or chymotrypsin), particularly in fish fed on low FM diet. Trypsin activity displayed a different response to feeding frequency between the different diets, with a significant decreasing effect at an increasing number of meals observed only under low FM-FO dietary level. Tian et al. (2015) did not find significant differences in protease activity at increasing feeding frequency in snout bream juvenile, while Xie et al. (2011) found that the intestinal trypsin activity increased at increasing feeding frequencies up to 8 meals per day$^{-1}$, which was considered the optimal frequency for large yellow croaker larvae. According to Tian et al. (2015), an optimal feeding frequency might enhance the intestinal digestive function of fish, and increased activity of a digestive enzyme generally indicates enhanced ability of an animal to obtain nutrients from feed. In the present study, no significant differences in trypsin activity were detected when comparing different diets after one single meal (FM30/FO15$_{F1}$ *vs* FM10/FO3$_{F1}$) indicating that the reduction in trypsin activity observed at increasing feeding frequency only under low FM-FO dietary level resulted from the interaction of diet and feeding frequency. FM10FO3 diet is formulated based on vegetable ingredients and the presence of residual protease inhibitors in vegetable aquafeed ingredients is known to affect digestive enzyme activity, as previously reported in sea bream and other fish species (Krogdahl et al., 2003; Santigosa et al., 2008; Yaghoubi et al., 2016; Parma et al., 2019; Parma et al., 2020); and according to Moyano et al. (1999) it depends on the type and amount of plant meal extension of the feeding period and sensitivity of fish species. A reduction in both storage and production of the digestive enzymes in the hepatopancreas could also be responsible for a reduction in the activity after consecutive meals as recently reported in the Sparids red sea bream (*Pagrus major*) fed on long-term high soybean meal dietary level (Murashita et al., 2018). Further studies are necessary to clarify the role of vegetable ingredients and associated feeding frequency affecting digestive enzyme activity; however, in the present study it seemed that consecutive meals could have amplified the potential trypsin inhibitor effect of the vegetable meal-based diet adopted. Nevertheless, fish receiving several meals presented similar values of growth, feed conversion or protein digestibility to those fed only once.

This can be explained when considering the estimated values of total enzyme production per day; in this case, the secretion of trypsin associated with repeated meals could compensate the lower time available for hydrolysis derived from faster gut transit rates. The net result should be the observed similarity in total performance and use of nutrients by all fish groups. Nevertheless, this compensation was not equally achieved in fish fed on low FM diets for some of the above-indicated reasons. This could have contributed to determining the lower values of FCR and growth measured in these fish, particularly in those fed three times a day. In relation to pH, according to Yufera et al. (2012, 2014), in Sparids the hydrochloric acid is secreted just after the ingestion of food and activates the transformation of pepsinogen to pepsin when gastric pH decreases below 4.0. Montoya et al. (2010a) have shown a drop in pH to 4 at 4 h postprandial in gilthead sea bream (83±4.8g), while in the same species Nikolopoulou et al. (2011) observed a decline of gastric pH after one meal, from 5.5 to 2.5 within 8 h, and also that restoration of neutral gastric pH (which indicates the completion of gastric digestion) required a longer time. However, Yúfera et al. (2014) found that in early juveniles (5-7g) of this species, the minimum gastric pH changed from 4 to 12 hours post-feeding depending on feeding frequency. In the present study no measurements of gut pH were carried out, but it is presumed that increasing the number of meals precluded achieving an optimal gastric pH for pepsin. Nevertheless, as previously indicated, this seemed to be compensated by the net increased secretion of enzyme associated with repeated feeding as well as by a more favourable enzyme:substrate ratio during hydrolysis.

    However it must be remembered, that daily patterns of ingestion, gastric acidification and enzymatic activity could be different in fish fed a fixed daily ration vs those fed to satiation at each meal. In addition, the reduction of dietary FM level in favour of vegetable

ingredients may determine a reduction in the feed buffering capacity with possible effects on the pattern of gastric acidification (Marquez et al., 2012). It is possible that a lower buffering capacity of FM10FO3 in comparison to FM30FO15 may have determined a faster drop in gastric pH, with consequences on the timing of pepsin activation after food ingestion.

Amylase activity displayed a similar response to feeding frequency in both dietary treatments characterised by an increasing pattern from 1 to 2 meals a day$^{-1}$ followed by a reduction when 3 meals a day$^{-1}$ were offered. Amylase activities displayed different responses to feeding regime and daily rhythms in different fish species. In agreement with our findings, the juvenile blunt snout bream (*Megalobrama amblycephala*) showed an increase in α-amylase activity as daily feeding frequency increased from 1 to 3 times, but a further increase in feeding frequency led to a decrease in activity (Tian et al., 2015). In white sea bream (*Diplodus sargus*), amylase activity was higher in fish fed twice a day than in fish fed 3 or 4 times a day (Enes et al., 2015) while in tambaqui (*Colossoma macropomum*) amylase activity followed a clear daily rhythm irrespective of feed intake (Reis et al., 2019). Similarly, in the Nile Tilapia (*Oreochromis niloticus*) amylase activity remains constant during the day irrespective of feeding frequency, indicating that the omnivorous feeding habit of this species required constant activity of this enzyme (Guerra-Santos et al., 2017). Interestingly, in gilthead sea bream, Montoya et al. (2010a) found a feeding synchronisation rhythm in the amylase activity, which increased a few hours before mealtimes only when fish were fed at fixed feeding times. The same authors also found that the amylase activity level remains stable within a few hours after feeding. In the present study, the tendency of increasing amylase activity values observed under two daily meals could be related to a possible increased in the activity achieved prior to the second meal. Otherwise, it is likely that the short interval between meals in frequency 3 led to a lower and constant level of activity. No significant effects of feeding frequency were observed for chymotrypsin and lipase activity indicating that the activity of these enzymes was independent from feeding delivery times. Few studies reported the relation among these enzymes with feeding procedures. Recently, in juvenile arapaima (*Arapaima gigas*) no significant differences were observed in lipase activity following different feeding strategies including a self-feeding system, 2 and 3 daily meals to satiation, and feeding at fixed ration (Pedrosa et al., 2019). The authors suggest that, despite requiring extra energy expenditure, keeping the full set of digestive enzymes available in that species is a helpful strategy to utilize all available food found in the environment (Pedrosa et al., 2019).

The study of plasma biochemistry has been widely employed in relation to acute or chronic stressors and metabolic disorders, or to investigate the effects of different feed ingredients, feeding practices and rearing conditions (Peres et al., 2013; Guardiola et al., 2018; Bonvini et al., 2018a, b). In the present study, although feeding frequency did not affect most of the plasma parameters analysed, it seemed to be more relevant than diet in affecting plasma parameters. In general, overall plasma values were consistent with reported data on juveniles and on-growing sea bream under optimal nutritional conditions (Peres et al., 2013; Parma et al., 2020). In common with the present study, it was found that different feeding frequency did not influence glycaemia levels in white sea bream 4 h after feeding (Enes et al., 2015), as also reported for glucose, total protein and triacylglycerol in other fish species fed different feeding ratio or frequency (Cho et al., 2007; Guo et al., 2018; Pedrosa et al., 2019). That glucose levels are independent of the amount of food and the time of administration, as observed in the present study, is in line with previous observation confirming that sea bream is able to restore basal glucose levels relatively quickly after feeding (Peres et al., 2013). The same authors observed a reduction in triglycerides due to mobilisation of body fat reserves only during the first week of starvation while the decrease in cholesterol and total protein begin after a week of fasting. Plasma levels of non-specific enzymes such as AST, ALT, ALP, CK and LDH may provide important information on the health status of fish; an elevated level may indicate specific tissue damage in several organs including liver, muscle, spleen and kidney (Peres et al., 2013; Guardiola et al., 2018). However, few studies in fish are available in relation to feeding frequency. Specifically, an increase in AST and ALT has been correlated with an increase in feeding frequency, indicating a possible cause of hepatic damage and alteration of triglycerides due to inappropriate feeding frequency (Guo et al., 2017), while an increase in ALT and ALP activity in fish fed twice a day compared to those fed 3 times was related with the larger amount of food delivered in a short time (Pedrosa et al., 2019). The absence of differences in most of the non-specific plasma enzymes of the present study suggested no major functional changes or suffering of organs responsible for metabolism and catabolism, such as liver and kidney, at increasing feeding frequency. However, plasma creatinine displayed higher values in F1 compared to F2 and F3 under high FM-FO dietary level. Plasma creatinine is a routine marker of kidney function in humans and has also been employed to study kidney toxicity related to selenium dietary levels in Atlantic salmon (*Salmo salar*) (Gowda et al., 2010; Berntssen et al., 2018). No reference values for gilthead sea bream are available; however, the creatinine values reported in FM30FO15$_{F1}$ (0.64 ± 0.29 mg dL$^{-1}$) were higher than those (0.12-0.37 mg dL$^{-1}$) reported in other studies of the same species and size fed to satiation twice a day using similar diets (Parma et al., 2020), or fed standard or low FM-FO dietary levels with 1-2 daily meals to satiation (Benedito-Palos et al., 2016). The higher values observed seem to depended on both feeding and dietary regime as highlighted by a significant interactionThus, the combination of large amounts of food ingested in a short time, together with high FMFO diet (which also provided a higher protein utilisation in comparison to low FMFO level), could have led to a physiological increase in plasma creatinine as a result of faster protein metabolism under this diet and feeding regime. Among electrolytes, feeding frequency also showed an effect on plasma $Na^+$ and Cl with increasing values observed at higher feeding frequencies (F2 and F3). This effect could be related to the different amount of seawater ingested. Infact the water necessary for feed moisturization in the gastrointestinal tract derives from drinking and stomach secretions (Bonvini et al., 2018a). In addition, a significant interaction indicated that the combined effect of frequency and diet led to a lower Na/K in F3 and F1 for FM30/FO15 and FM10/FO3, respectively. Cortisol level was affected by feeding frequency, with higher values recorded under two daily meals. Previous studies have highlighted the role of mealtime, feeding frequency and feeding behaviour as the synchroniser and modulator of plasma cortisol in sea bream and other fish species, as reviewed by Lopez-Olmeda et al. (2012). In sea bream fed a single daily meal during daylight hours the elevation in cortisol levels was found 2h after the meal, while the recovery of basal values took 4 h after the meal (Montoya et al. 2010b). Accordingly, the differences reported in the present study seem to be determined by the greater time occurring between the two meals when two daily meals were offered during daylight hours. However, it should also be mentioned that the amount of feed supplied per meal in the feeding frequencies was different, although distributed during the same overall meal time, and this may have caused some stress to the fish trying to reach pellets during feeding in F2 and F3.

In conclusion, the different feeding frequencies tested in this trial had no major effects on overall performances and feed efficiency during the on-growing of gilthead sea bream fed high or low fishmeal and fish oil dietary level. Although the activity of pepsin and trypsin responded differently in relation to feeding frequency and diets, the cumulative process of digestion and nutrient uptake as expressed by apparent nutrient digestibility and nutritional indices seemed equally efficient regardless of the number of meals in both high and low marine-based diets. The absence of a significant interaction indicates that the combined effect of frequency and diet did not alter the growth and nutritional indices studied and that feeding frequency did not amplify or reduce the differences observed between diets. Most of the plasma parameters related to nutritional and physiological conditions were not affected by feeding frequency and were in line with normal values for optimal healthy condition in this species. However, while FMFO level was the major factor affecting growth and nutritional indices, feeding frequency was the major factor affecting some plasma parameters such as Na, Cl, cortisol and creatinine. The higher level of plasma creatinine detected in fish fed a single daily meal with high FMFO level seems to be within physiological values in relation to the high protein efficiency observed with this diet. However, further study is required in order to exclude a possible overload of kidney functionality. According to the results, gilthead sea bream seems able to maximize feed utilisation regardless of the number of meals using high or low FM-FO dietary level, and this could be a useful indicator when planning feeding activity at farm level in order to optimise fish growth and costs of feeding procedures.

**Declaration of Competing Interest**

The authors declare that they have no known competing financial interests or personal relationships that could have appeared to influence the work reported in this paper.

**Acknowledgments**

This research was undertaken under the MedAID (Mediterranean Aquaculture Integrated Development) project, which has received funding from the European Union's Horizon 2020 Research and Innovation Programme, Call H2020-SFS-23-2016, Grant agreement no 727315 (http://www.medaid-h2020.eu/). The authors would like to thank Gillian Forlivesi Heywood for English language editing.

**Table 1.** Ingredients and proximate composition of the experimental diets

|  | FM30/FO15 | FM10/FO3 |
|---|---|---|
| *Ingredients, % of the diet* | | |
| Fish meal (LT70) | 30.0 | 10.0 |
| Soybean meal 48 | 9.0 | 9.0 |
| Soy protein concentrate | 10.0 | 20.5 |
| Wheat gluten | 5.0 | 10.2 |
| Corn gluten | 10.0 | 15.0 |
| Wheat meal | 9.7 | 7.3 |
| Rapeseed meal | 5.0 | 4.0 |
| Sunflower meal | 5.0 | 4.0 |
| Fish oil | 15.0 | 3.0 |
| Rapeseed oil | 0 | 13.0 |
| Vit/Min premix[1] | 1.0 | 1.0 |
| Antioxidant powder (Paramega) | 0.2 | 0.2 |
| Sodium propionate | 0.1 | 0.1 |
| MCP |  | 2.0 |
| Lysine | - | 0.3 |
| Methionine | - | 0.1 |
| L-Tryptophan |  | 0.3 |
| *Proximate composition, % on a wet weight basis* | | |
| Moisture | 5.1 | 4.8 |
| Protein | 45.6 | 46.7 |
| Lipid | 16.8 | 16.9 |
| Ash | 7.9 | 6.4 |
| Gross energy cal g$^{-1}$ | 5043.6 | 5019.38 |

[1]Vitamins and mineral premix (IU or mg kg-$^{1}$ diet; Invivo NSA,: Portugal); DL-alpha tocopherol acetate, 200 mg; sodium menadione bisulphate, 10 mg; retinyl acetate, 16650 IU; DL-cholecalciferol, 2000 IU; thiamine, 25 mg; riboflavin, 25 mg; pyridoxine, 25 mg; cyanocobalamin, 0.1 mg; niacin, 150 mg; folic acid, 15 mg; L-ascorbic acid monophosphate, 750 mg; inositol, 500 mg; biotin, 0.75 mg; calcium panthotenate, 100 mg; choline chloride, 1000 mg, betaine, 500 mg; copper sulphate heptahydrate, 25 mg; ferric sulphate monohydrate, 100 mg; potassium iodide, 2 mg; manganese sulphate monohydrate, 100 mg; sodium selenite, 0.05 mg; zinc sulphate monohydrate, 200 mg; Yttrium oxide, 100 mg
MCP: monocalcium phosphate

**Table 2.** Growth performance of gilthead sea bream fed diets with low and high FM and FO level under different feeding frequencies (F) over 109 days.

| | *Experimental diets* | | | | | | | *P-value* | |
|---|---|---|---|---|---|---|---|---|---|
| | FM30/FO15 | | | FM10/FO3 | | | Frequency | Diet | Interaction |
| | 1F | 2F | 3F | 1F | 2F | 3F | | | |
| IBW (g) | 86.3±2.2 | 89.7±3.3 | 87.7±1.3 | 89.5±2.1 | 87.7±1.0 | 89.0±3.6 | *0.860* | *0.469* | *0.218* |
| FBW (g) | 273.2±12.7 | 279.0±8.2 | 275.2±5.6 | 261.6±3.5 | 262.0±3.4 | 260.2±5.6 | *0.729* | *0.001* | *0.807* |
| Weight gain (g) | 186.9±10.5 | 189.3±6.1 | 187.5±4.5 | 172.1±1.9 | 174.2±3.8 | 171.2±8.5 | *0.780* | *0.000* | *0.980* |
| SGR | 1.06±0.02$^b$ | 1.04±0.02$^{ab}$ | 1.05±0.01$^{ab}$ | 0.98±0.01$^a$ | 1.00±0.02$^{ab}$ | 0.98±0.05$^a$ | *0.937* | *0.001* | *0.453* |
| FI | 12.3±0.22$^{ab}$ | 11.9±0.20$^a$ | 12.3±0.13$^{ab}$ | 12.6±0.22$^{ab}$ | 12.8±0.19$^b$ | 12.7±0.51$^{ab}$ | *0.200* | *0.002* | *0.638* |
| FCR | 1.33±0.03$^a$ | 1.30±0.00$^a$ | 1.33±0.02$^a$ | 1.41±0.02$^b$ | 1.44±0.03$^b$ | 1.45±0.05$^b$ | *0.433* | *0.005* | *0.495* |
| Survival % | 95.6±4.2 | 96.1±1.9 | 93.9±4.8 | 97.8±1.0 | 93.9±1.0 | 95.0±1.7 | *0.404* | *0.797* | *0.406* |

Data are given as the mean (n=3) ± SD. In each line, different superscript letters indicate significant differences among treatments (Two-way Anova $p \leq .05$).
FM30/FO15 = diet formulated with 300g kg$^{-1}$ fishmeal (FM), 150 g kg$^{-1}$ fish oil (FO); FM10/FO3 = diet formulated with 100g kg$^{-1}$ FM; 30g kg$^{-1}$ FO. F1, one meal day$^{-1}$; F2, 2 meals day$^{-1}$, F3, 3 meals day$^{-1}$. IBW = Initial body weight.
FBW = Final body weight.
SGR = Specific growth rate (% day$^{-1}$) = 100 * (ln FBW- ln IBW) / days.
FI = Feed intake (g kg ABW$^{-1}$ day$^{-1}$) = ((1000*total ingestion)/(ABW))/days)).
FCR = Feed conversion rate = feed intake (g) / weight gain (g).

**Table 3**. Body composition and nutritional indices of gilthead sea bream fed diets with low and high FM and FO level under different feeding frequencies (F) over 109 days.

|  | *Experimental diets* | | | | | | *P-value* | | |
|---|---|---|---|---|---|---|---|---|---|
|  | FM30/FO15 | | | FM10/FO3 | | | | | |
|  | F1 | F2 | F3 | F1 | F2 | F3 | Frequency | Diet | Inter. |
| *Whole body composition*, % | | | | | | | | | |
| Protein | 17.7 ± 0.1$^{ab}$ | 17.8 ± 0.3$^b$ | 17.8 ± 0.3$^b$ | 17.3 ± 0.1$^a$ | 17.4 ± 0.2$^{ab}$ | 17.5 ± 0.1$^{ab}$ | 0.179 | 0.002 | 0.642 |
| Lipid | 16.3 ± 0.9 | 15.3 ± 0.1 | 15.3±0.6 | 15.5 ± 0.8 | 14.3 ± 0.2 | 16.4 ± 1.7 | 0.724 | 0.550 | 0.180 |
| Ash | 4.1 ± 0.2 | 3.7 ± 0.4 | 3.5±0.1 | 3.4 ± 0.4 | 3.9 ± 0.6 | 4.1±0.4 | 0.640 | 0.352 | 0.043 |
| Moisture | 61.2±1.1 | 62.1±1.0 | 61.6±1.4 | 63.2±0.7 | 62.9 ± 1.3 | 61.2±1.3 | 0.555 | 0.193 | 0.167 |
| *Nutritional indices* | | | | | | | | | |
| PER | 1.70 ± 0.05$^b$ | 1.73 ± 0.05$^b$ | 1.69 ± 0.01$^b$ | 1.53 ± 0.04$^a$ | 1.54 ± 0.05$^a$ | 1.52 ± 0.07$^a$ | 0.605 | 0.000 | 0.856 |
| Nutrient gain (g kg$^{-1}$ day$^{-1}$) | | | | | | | | | |
| DM | 3.91 ± 0.13 | 3.92 ± 0.16 | 3.85±0.14 | 3.55 ± 0.08 | 3.57 ± 0.15 | 3.78 ± 0.33 | 0.707 | 0.011 | 0.305 |
| Protein | 1.68 ± 0.02$^{bc}$ | 1.70 ± 0.02$^c$ | 1.71 ± 0.05$^c$ | 1.51 ± 0.05$^a$ | 1.54 ± 0.05$^a$ | 1.56 ± 0.07$^{ab}$ | 0.30 | 0.000 | 0.956 |
| Lipid | 1.80 ± 0.11 | 1.76 ± 0.12 | 1.69 ± 0.05 | 1.58 ± 0.18 | 1.60 ± 0.14 | 1.79 ± 0.28 | 0.783 | 0.230 | 0.216 |
| Nutrient retention (%) | | | | | | | | | |
| DM | 33.4 ± 1.0$^{ab}$ | 34.6 ± 1.5$^b$ | 33.0 ± 1.5$^{ab}$ | 29.6 ± 0.3$^a$ | 29.4 ± 1.6$^a$ | 31.3 ± 2.6$^{ab}$ | 0.632 | 0.001 | 0.163 |
| Protein | 29.9 ± 0.9$^b$ | 31.3 ± 0.6$^b$ | 30.6 ± 0.8$^b$ | 25.7 ± 1.2$^a$ | 25.9 ± 0.9$^a$ | 26.3 ± 1.4$^a$ | 0.366 | 0.000 | 0.514 |
| Lipid | 87.3 ± 4.5 | 88.3 ± 6.8 | 82.3 ± 3.5 | 74.3 ± 8.6 | 74.2 ± 7.5 | 83.5 ± 13.1 | 0.891 | 0.041 | 0.219 |
| *Morphological parameters*, % | | | | | | | | | |
| HSI | 1.69 ± 0.13 | 1.63 ± 0.12 | 1.64 ± 0.15 | 1.80 ± 0.09 | 1.60 ± 0.14 | 1.81 ± 0.13 | 0.326 | 0.262 | 0.519 |
| VSI | 6.57 ± 0.43 | 6.11 ± 0.26 | 5.98 ± 0.27 | 6.39 ± 0.25 | 6.38 ± 0.32 | 6.70 ± 0.69 | 0.545 | 0.129 | 0.117 |
| MFI | 1.88 ± 0.26 | 1.54 ± 0.12 | 1.44 ± 0.27 | 1.11 ± 0.33 | 1.29 ± 0.30 | 1.11 ± 0.08 | 0.344 | 0.000 | 0.275 |

Data are given as the mean (n=3) ± SD. HSI, VSI, MFI n=15 ± SD. In each line, different superscript letters indicate significant differences among treatments (Two-way Anova $p \leq .05$). FM30/FO15 = diet formulated with 300 g kg$^{-1}$ fishmeal (FM), 150 g kg$^{-1}$ fish oil (FO); FM10/FO3 = diet formulated with 100 g kg$^{-1}$ FM; 30 g kg$^{-1}$ FO. F1, one meal day$^{-1}$; F2, 2 meals day$^{-1}$, F3, 3 meals day$^{-1}$.
PER = Protein efficiency ratio = ((FBW-IBW)/protein intake).
Nutrient gain (g kg$^{-1}$ day$^{-1}$) = (final carcass dry matter DM, protein or lipid content, g - initial carcass DM, protein or lipid content, g) / ABW, kg / days.
Nutrient retention (%) = DM, protein or lipid gain / DM, protein or lipid intake (where DM, protein or lipid intake = nutrient intake, g / ABW, kg / days).
HSI = Hepatosomatic index (%) = 100 ∗ (liver weight / FBW).
VSI = Viscerosomatic index (%) = 100 ∗ (viscera weight / FBW).
MFI = Mesenteric fat index (%) = 100 ∗ (mesenteric fat weight / FBW).

0

**Table 4.** Feed digestibility of gilthead sea bream fed diets with low and high FM and FO level under different feeding frequencies (F)

| | | *Experimental diets* | | | | | | *P-value* | |
|---|---|---|---|---|---|---|---|---|---|
| | | FM30/FO15 | | | FM10/FO3 | | | | |
| | F1 | F2 | F3 | F1 | F2 | F3 | *Frequency* | *Diet* | *Inter.* |
| Dry matter | 95.1 ± 0.5 | 95.8 ± 0.9 | 96.3 ± 1.0 | 96.2 ± 0.3 | 96.6 ± 0.0 | 96.7 ± 0.9 | *0.124* | *0.037* | *0.723* |
| Protein | 74.1 ± 3.1 | 75.3 ± 5.6 | 78.1 ± 5.1 | 79.1 ± 1.0 | 81.0 ± 1.1 | 80.3 ± 4.9 | *0.538* | *0.038* | *0.721* |
| Energy | 67.1 ± 2.8 | 65.7 ± 7.7 | 69.7 ± 7.3 | 68.2 ± 1.3 | 68.8 ± 2.1 | 66.6 ± 8.3 | *0.967* | *0.888* | *0.646* |

Data are given as the mean (n = 3) ± SD. FM30/FO15 = diet formulated with 300g kg$^{-1}$ fishmeal (FM), 150 g kg$^{-1}$ fish oil (FO); FM10/FO3 = diet formulated with 100g kg$^{-1}$ FM; 30g kg$^{-1}$ FO. F1, one meal day$^{-1}$; F2, 2 meals day$^{-1}$, F3, 3 meals day$^{-1}$.

**Table 5.** Plasma biochemistry values for sea bream fed diets with low and high FM and FO level under different feeding frequencies (F) over 109 days.

| | *Experimental diets* | | | | | | *P-value* | | |
|---|---|---|---|---|---|---|---|---|---|
| | FM30/FO15 | | | FM10/FO3 | | | | | |
| Parameters | F1 | F2 | F3 | F1 | F2 | F3 | *Frequency* | *Diet* | *Inter.* |
| Glucose (mg dL$^{-1}$) | 113±27 | 135±30 | 113±20 | 101±28 | 123±23 | 115±30 | *0.063* | *0.286* | *0.615* |
| Urea (mg dL$^{-1}$) | 10.8±2.5 | 11.1±2.2 | 12.6±3.0 | 11.6±3.1 | 10.5±1.3 | 10.4±3.1 | *0.769* | *0.390* | *0.211* |
| Creatinine (mg dL$^{-1}$) | 0.64±0.29[b] | 0.22±0.17[a] | 0.22±0.07[a] | 0.43±0.22[ab] | 0.49±0.24[ab] | 0.31±0.25[ab] | *0.008* | *0.505* | *0.026* |
| Uric acid (mg dL$^{-1}$) | 0.87±0.54 | 0.68±0.53 | 0.39±0.27 | 0.63±0.50 | 0.51±0.45 | 0.45±0.39 | *0.108* | *0.399* | *0.603* |
| Tot bil (mg dL$^{-1}$) | 0.17±0.11 | 0.11±0.10 | 0.11±0.05 | 0.19±0.11 | 0.23±0.14 | 0.13±0.12 | *0.273* | *0.098* | *0.362* |
| CHOL (mg dL$^{-1}$) | 286±43 | 204±101 | 184±19 | 229±111 | 255±180 | 234±47 | *0.291* | *0.602* | *0.153* |
| HDL | 53.1±24.1 | 50.7±16.1 | 36.4±8.0 | 47.4±19.9 | 76.6±39.0 | 49.3±14.9 | *0.065* | *0.104* | *0.153* |
| TRIG (mg dL$^{-1}$) | 847±340 | 808±290 | 797±162 | 750±272 | 701±453 | *1016*±331 | *0.479* | *0.961* | *0.339* |
| TP (mg dL$^{-1}$) | 3.85±0.43 | 3.87±0.42 | 3.73±0.31 | 3.90±0.58 | 4.29±1.48 | 4.17±0.29 | *0.704* | *0.122* | *0.604* |
| ALB (g dL$^{-1}$) | 0.93±0.11 | 0.92±0.08 | 0.84±0.32 | 0.95±0.16 | 0.94±0.37 | 1.04±0.06 | *0.991* | *0.209* | *0.385* |
| AST (U L$^{-1}$) | 13.0±8.8 | 13.0±7.8 | 11.3±11.8 | 9.3±3.6 | 10.0±9.0 | 15.6±7.6 | *0.772* | *0.797* | *0.458* |
| ALP (U L$^{-1}$) | 409±132 | / | 480±215 | 512±249 | / | 538±158 | *0.472* | *0.243* | *0.793* |
| CK (U L$^{-1}$) | 8.08±4.70 | 4.17±2.86 | 3.11±2.62 | 5.70±6.06 | 2.80±2.17 | 8.33±8.36 | *0.189* | *0.744* | *0.076* |
| LDH (U L$^{-1}$) | 469±498 | 362±226 | 318±239 | 424±408 | 304±208 | 394±312 | *0.725* | *0.946* | *0.898* |
| Ca$^{+2}$ (mg dL$^{-1}$) | 14.3±1.0 | 14.9±0.7 | 13.6±0.6 | 14.7±2.1 | 14.9±3.6 | 14.8±1.0 | *0.625* | *0.295* | *0.669* |
| P (mg dL$^{-1}$) | 14.7±3.1 | 14.4±2.2 | 12.3±0.7 | 15.0±3.7 | 14.9±5.1 | 14.3±1.9 | *0.313* | *0.324* | *0.696* |
| K$^+$ (mEq L$^{-1}$) | 7.21±3.18 | 6.15±1.52 | 8.33±1.71 | 8.02±2.21 | 6.49±0.98 | 6.45±1.28 | *0.193* | *0.671* | *0.086* |
| Na$^+$ (mEq L$^{-1}$) | 191±7[a] | 211±7[b] | 204±12[ab] | 191±17[a] | 210±4[b] | 210±12[b] | *0.000* | *0.566* | *0.632* |
| Fe (μg dL$^{-1}$) | 169±56 | 138±31 | 132±34 | 168±48 | 203±79 | 167±32 | *0.427* | *0.031* | *0.201* |
| Cl (mEq L$^{-1}$) | 152±4[a] | 168±7[b] | 165±10[b] | 152±15[a] | 161±5[ab] | 167±8[b] | *0.000* | *0.518* | *0.363* |
| Mg (mg dL$^{-1}$) | 4.08±0.92 | 3.85±0.44 | 3.08±0.52 | 3.84±0.80 | 4.09±1.20 | 3.58±0.70 | *0.053* | *0.492* | *0.386* |
| Cortisol (μg dL$^{-1}$) | 1.63±0.84[a] | 5.11±0.81[ab] | 2.16±0.98[a] | 3.97±1.44[ab] | 8.03±8.45[b] | 2.84±0.49[ab] | *0.008* | *0.052* | *0.614* |
| ALB/GLOB | 0.32±0.03 | 0.31±0.01 | 0.31±0.11 | 0.32±0.02 | 0.29±0.02 | 0.34±0.03 | *0.701* | *0.840* | *0.582* |
| CaxP | 212±51 | 213±31 | 166±9 | 211±61 | 216±122 | 213±38 | *0.442* | *0.362* | *0.425* |
| Na/K | 31.5±13.2 | 35.8±7.5 | 25.5±5.0 | 25.5±7.1 | 33.1±6.0 | 34.0±8.1 | *0.129* | *0.978* | *0.017* |

Data are given as the mean ± SD. Different letters indicate significant difference (Two-way Anova $p ≤ .05$) between treatments. FM30/FO15 = diet formulated with 300g kg$^{-1}$ fishmeal (FM), 150 g kg$^{-1}$ fish oil (FO); FM10/FO3 = diet formulated with 100g kg$^{-1}$ FM; 30g kg$^{-1}$ FO. F1, one meal day$^{-1}$; F2, 2 meals day$^{-1}$, F3, 3 meals day$^{-1}$. Tot Bil, total bilirubin; CHOL, cholesterol; HDL, high density lipoprotein; TRIG, triglycerides; TP, total protein; ALB, albumin; AST, aspartate aminotransferase; ALP, alkaline phosphatase; CK, creatine kinase; LDH, lactate dehydrogenase, Ca$^{+2}$, calcium; P, inorganic phosphorus; K$^+$, potassium; Na$^+$, sodium; Fe, iron; Cl, chloride; Mg, magnesium; GLOB, globuline.

## Key to Figures

Figure 1. (A) pepsin activity (units g live weight$^{-1}$) per meal and (a) daily estimated; (B) trypsin activity (units g live weight$^{-1}$) per meal and (b) daily estimated; (C) chymotrypsin activity (units g live weight$^{-1}$) per meal and (c) daily estimated; (D) amylase activity (units g live weight$^{-1}$) per meal and (d) daily estimated; (E) lipase activity (units g live weight$^{-1}$) per meal and (e) daily estimated; F trypsin/chymotrypsin per meal, (N=9) ± SD measured in the gastrointestinal tract of gilthead sea bream fed high and low FM-FO dietary level under different feeding frequency (F).  FM30/FO15 = 300g kg$^{-1}$ fishmeal (FM), 150 g kg$^{-1}$ fish oil (FO);  FM10/FO3 = 100g kg$^{-1}$ FM; 30g kg$^{-1}$ FO. F1, one meal day$^{-1}$; F2, 2 meals day$^{-1}$, F3, 3 meals day$^{-1}$. Different letters indicate significant differences between treatments (Two-way ANOVA, $p < .05$; F = feeding frequency effect; D = diet effect; I = interaction).

Pepsin

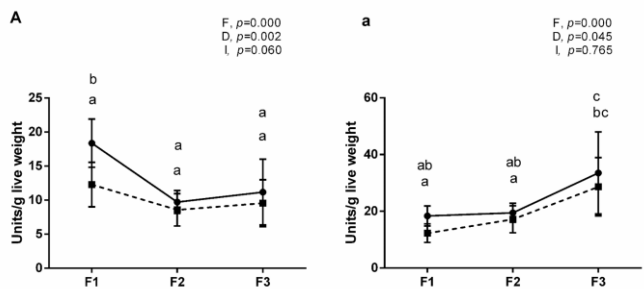

Trypsin

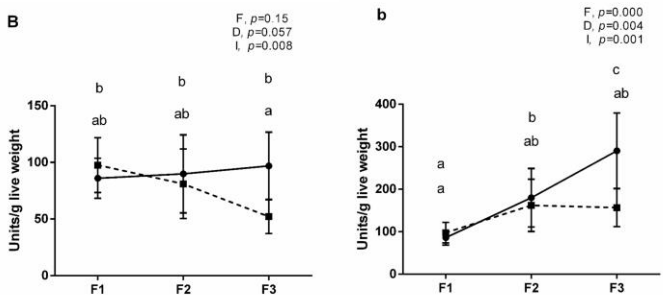

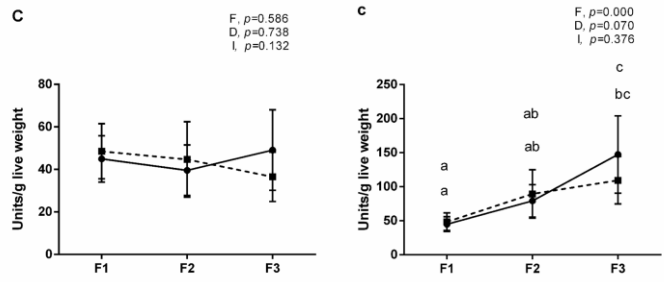

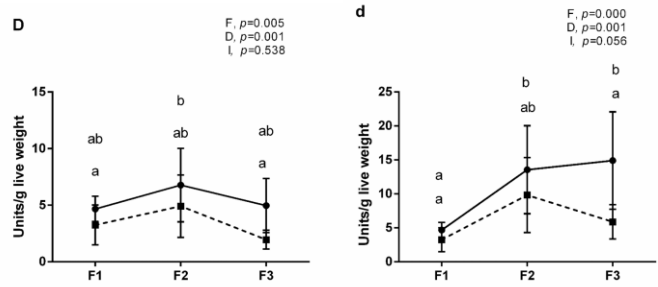

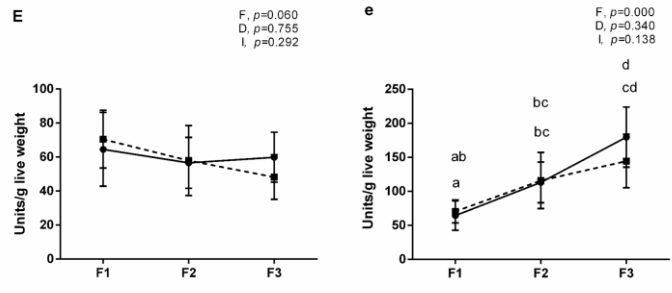

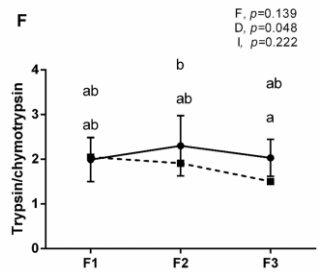